\documentstyle[preprint,aps,amsfonts]{revtex}
\def\vec#1{{\bf#1}}

\begin{document}
\draft
\title{Relation between Barrier Conductance and Coulomb Blockade
Peak Splitting for Tunnel-Coupled Quantum Dots}
\author{John M. Golden and Bertrand I. Halperin}
\address{Department of Physics, Harvard University, Cambridge, MA
  02138}

\date{Submitted to Phys. Rev. B, 2 May 1995}

\maketitle

\begin{abstract}
\noindent \ \ We study the relation between the barrier conductance
and the Coulomb blockade peak splitting for two
electrostatically equivalent dots connected by tunneling
channels with bandwidths much larger than the dot charging
energies.  We note that this problem is
equivalent to a well-known single-dot problem and present
solutions for the relation between peak splitting and barrier
conductance in both the weak and strong coupling limits.  Results
are in good qualitative agreement with the experimental findings
of F.~R.~Waugh~et~al.
\end{abstract}



\narrowtext

\section{Introduction}

Turning on a tunnel junction to a quantum dot leads to
progressive destruction of the single-dot Coulomb blockade.
Experiments by Waugh~et~al.~\cite{Waugh} and
Molenkamp~et~al.~\cite{Molen} chronicle this eradication
for two tunnel-coupled dots of equal and widely
disparate charging energies, respectively.
Inspired by the experimental results of Ref.~\cite{Waugh},
the present paper seeks to develop a
simple model for the coherent tunneling of electrons
between a pair of electrostatically identical quantum dots.
(See Figure~1(a) for a schematic view of Waugh~et~al.'s
double-dot structure.)
The goal is to describe the evolution of the Coulomb
blockade from that of two isolated dots to that of one
composite dot in terms of parameters that determine the
states of the isolated dots and the nature of the
connection between them.  In the limits relevant to the
experimental situation in Ref.~\cite{Waugh}, we find that the
most important dimensionless parameters are the number
$N_{ch}$ of conducting channels between the two dots
and the dimensionless {\em interdot barrier conductance} $g$
of each channel, which is measured when the
Coulomb blockade has been removed.
(The interdot barrier conductance was measured in
Ref.~\cite{Waugh} by de-energizing the external barrier
potentials $V_{xi}$
that separate the dots from the leads.  This conductance is to
be distinguished from the conductance measured in
the double-dot Coulomb blockade measurements, which will
be referred to as the {\em Coulomb blockade conductance} or
{\em double-dot conductance}.)

The problem of coupled quantum dots and more generally, of the
effect of tunnel-couplings upon the Coulomb blockade has
received much attention.  I.~M.~Ruzin~et~al.~\cite{Ruzin} examined the
Coulomb blockade structure of two non-identical dots in series
via a standard activation-energy approach.  C.~A.~Stafford
and S.~Das~Sarma~\cite{Stafford1,Stafford2} as well
as G.~Klimeck~et~al.~\cite{Klimeck} have
applied Hubbard-like models with and without interdot
capacitances to determine the many-body wavefunctions for
tunnel-coupling between a small array of single-dot eigenstates.
Many investigators have studied the effect of tunneling upon
the Coulomb blockade for metallic junctions, in which there
are a large number of conducting channels [5-9,17].  Relatively
few have considered junctions with only one or two
channels~[16,18-20].

In Section II of this paper, we present a brief review of the
experimental results that have motivated
our investigation.  In Section III, we define a tunneling
model which is useful for calculations in the limit of weak
coupling between the two dots.  The strong-coupling limit is
analyzed in Section IV, and a summary of our conclusions is
presented in Section V.

\section{Motivation}

The experiment of F.~R.~Waugh~et~al.~\cite{Waugh} provides
the primary motivation for this paper.  These authors study the
effect that varying the interdot potential barriers has
upon the Coulomb blockade conductance peak structure for arrays of
$n$ dots, where $n$ equals 2 or 3.  For their Coulomb blockade
measurements, they energize the confining gates ($V_{xi}$ in
Figure~1(a)) so that the conductance between the dots and
the external leads is much less than $2e^{2}/h$.
Having tuned the dots to be electrostatically
identical---i.e., to have common gate and total
dot capacitances $C_{g}$ and $C_{\Sigma}$---they find that
lowering the interdot barriers results in
interpolation between the peak structure characteristic of the
isolated individual dots and that characteristic of a single
composite dot having capacitance $nC_{\Sigma}$: the
initial isolated-dot peaks split into bunches of
$n$ sub-peaks, and the splitting within the bunched sub-peaks increases
until they are essentially equally distributed with $n$-times
the periodicity of the original peaks.  (See Figures~2(b) and 2(c).)
For the double dot ($n=2$),
Waugh~et~al.\ also measure the conductance $G_{b}$ of the barrier
between the two dots after the exterior walls of the double dot
have been removed.  In plots of the sub-peak splitting and barrier
conductance as functions of the barrier gate voltage, they remark
that the latter appears to be a displaced repetition of the former,
the displacement being explicable as a result of the
need to correct for the exterior walls'\ influence upon the
barrier.

Waugh~et~al.\ use a $T=0$ ``capacitive charging model'' to interpret
their data.  In this model, electrons on the dots are treated as
charged particles with no kinetic energy that occupy each
dot in integer amounts.  In the absence of coupling, the energy is
given by the sum of the potential energies of the individual dots.
For two dots with common capacitances $C_{\Sigma}$ and $C_{g}$, the
expression for the energy has a particularly simple form:
\begin{equation}
     E = \frac{U}{2} \sum_{i=1}^{2} (n_{i}-\phi_{i})^{2},
\label{eq:capchg}
\end{equation}
where $U$ is the charging energy for each individual dot,
$U=e^{2}/C_{\Sigma}$;
$n_{i}$ is the number of electrons on the $i$th dot; and
$\phi_{i}$ is the gate voltage parameter that determines the
energy-minimizing value of $n_{i}$.  For common gate voltages
and gate-to-dot capacitances, we have the relations
\mbox{$\phi_{i} \equiv C_{gi}V_{gi}/e = C_{g}V_{g}/e \equiv \phi$}.
Figure~1(a) should help put these parameters in context.

For each set of integer occupation numbers $(n_{1},n_{2})$,
the capacitive charging model with $\phi_{i}=\phi$
gives an energy $E_{(n_{1},n_{2})}$
that is a parabolic function of the common gate voltage parameter
$\phi$.  (See Figure~2.)
All the parabolas are identical in shape, their only
distinguishing features being the locations of their minima.
The lowest-energy parabola $E_{N_{tot}}(\phi)$ for a given
value of $N_{tot} = {\displaystyle \sum_{i=1}^{2} n_{i}}$
has $n_{1}=n_{2}=N_{tot}/2$ for $N_{tot}$ even and
$n_{1} = n_{2} \pm 1 = \frac{N_{tot} \pm 1}{2}$ for $N_{tot}$ odd.
In the former {\em even} case, the minima all lie on the line $E=0$.
In the latter {\em odd} case, the minima are displaced upward,
sitting along $E=U/4$.  For all parabolas, the $\phi$-coordinate
of the minimum is $N_{tot}/2$.

A prominent peak in the
double-dot conductance occurs at values of $\phi$ such that
the lowest-energy parabolas corresponding to consecutive values
of $N_{tot}$ cross---in other words, at values of $\phi$ for
which $E_{N_{tot}}(\phi)=E_{N_{tot}+1}(\phi)$ for some
integer $N_{tot}$.  For the model of
Equation~(\ref{eq:capchg}), this occurs whenever
$\phi=m+\frac{1}{2}$, where $m$ is an integer.  (One such crossing
point is marked by the black dot in Figure~2.)

In a model in which coupling between the dots is included,
the lowest-energy parabolas for odd $N_{tot}$ are
shifted downward relative to the lowest-energy even-$N_{tot}$
parabolas by an ``interaction energy'' $E_{int}$.  This downward
shift splits each of the initial crossing points into a pair
of crossing points symmetric about the position of the initial
degeneracy, from which they are separated by a distance
proportional to $E_{int}$.  As a result, each of the initial
conductance peaks is similarly split into two sub-peaks with
separation proportional to $E_{int}$.  The sub-peak splitting
reaches its saturation value when $E_{int}=U/4$---i.e., when
the lowest-energy {\em even} and {\em odd} parabolas
have the same minimum energy.  At this point, the relevant
crossing points occur for $\phi=\frac{1}{2} (m+\frac{1}{2})$.
The corresponding conductance peaks are once again equally
spaced, but their period is now that characteristic of
a single dot with capacitance $2 C_{\Sigma}$.

Thus, in the capacitive charging model, the problem of explaining
the peak splitting reduces to the problem of describing the shift
in the ground state energy of a double dot containing a fixed total
number of particles.  Waugh~\cite{Waugh} has shown that introduction
of a capacitive coupling $C_{int}$ between the two dots would allow
one to obtain a picture in qualitative agreement with the experimental
results: as the interdot capacitance goes to infinity, $E_{int}$
converges to $U/4$.  However, the magnitude of the interdot coupling
necessary to fit the experimental data is much larger than what one
would expect from an electrostatic interaction between two adjacent
dots having a narrow tunneling channel between them.  Consequently,
the use of the interdot capacitance $C_{int}$ must be regarded
as simply a reparametrization of the problem which replaces one
unknown, $E_{int}$, with another unknown, $C_{int}$.
What we really want is a theory which produces at least qualitative
agreement with experiment and expresses $E_{int}$
in terms of simple measurable quantities.  Waugh~et~al.\ provide
one candidate: the conductance $G_{b}$ of the barrier between
the two dots.  The remainder of this paper is devoted to
developing a theory of the relation between $E_{int}$ and
the dimensionless conductance per tunneling channel
$g=\frac{G_{b}}{N_{ch} G_{0}}$, where $N_{ch}$ is the number
of independent interdot tunneling channels (assumed to have
identical conductances) and $G_{0}$ is
the conductance quantum $e^{2}/h$.  In the experiment of
Ref.~\cite{Waugh}, there is no applied magnetic field and the
dots are connected by a narrow constriction allowing only a
single transverse orbital mode with double spin degeneracy.
As a result, in this experimental case, $N_{ch}$ equals 2.

\section{Tunneling Model for the Double-Dot Coupling}

\subsection{Definition of the Model}

Our goal can be stated a bit more
precisely.  For a general tunnel-coupling between two dots
involving any number $N_{ch}$ of identical, independent channels
and dimensionless channel conductance $g$, our aim is to
express the fractional energy shift $f \equiv \frac{E_{int}}{U/4}$
as a function of $g$ and $N_{ch}$ plus any other parameters that
might be found to be important.
In order to derive an equation for $f$, we
first choose a double-dot Hamiltonian.  Since we hope to
explain the evolution of the double-dot Coulomb blockade through
the barrier conductance alone, we will ignore---for the moment at
least---electrostatic coupling of the dots.  Interaction
between the dots will occur solely via tunneling
through the barrier between them.  Such tunnel Hamiltonians
have been found useful from the beginnings of Coulomb
blockade theory~\cite{Shekhter}, and the model we will use is
a double-dot version of the Hamiltonian used, for
example, by Averin and Likharev to investigate the conductance
oscillations of small metal-to-metal tunnel junctions~\cite{Averin1}.
In particular, we have the Hamiltonian $H=H_{0}+H_{T}$, where
\begin{eqnarray}
H_{0}  & = & K + V,  \nonumber \\
K_{\ } & = & \sum_{i=1}^{2} \sum_{\sigma} \sum_{\vec{k}}
            \epsilon_{i \vec{k} \sigma} \hat{n}_{i \vec{k} \sigma},
           \nonumber  \\
V_{\ } & = & \frac{U}{2} \sum_{i=1}^{2} (\hat{n}_{i} - \phi_{i})^{2},
             \nonumber  \\
H_{T}  & = & \sum_{\sigma} \sum_{\vec{k_1} \vec{k_2}}
         (t_{\vec{k_1} \vec{k_2}}
     c^{\dagger}_{2 \vec{k_2} \sigma}
  c_{1 \vec{k_1} \sigma} + h.c.).
\end{eqnarray}
In these equations, $i$ is the dot index, $\sigma$ is the channel
index (which could signify different spin channels), and $\vec{k}$
is the index for all internal degrees of freedom not included in
the channel index.  In addition,
$\hat{n}_{i}={\displaystyle \sum_{\vec{k} \sigma}} \hat{n}_{i \vec{k} \sigma}$
is the number operator for the $i$th dot, and
$t_{\vec{k_{1}} \vec{k_{2}}}$ is the tunneling matrix element between
a dot 1 wavefunction indexed by $\vec{k_1}$ and the dot 2 wavefunction
lying in the same channel and indexed by $\vec{k_2}$.  The gate voltage
parameter $\phi_{i}$ has the same meaning as in Equation~(\ref{eq:capchg}).
$\epsilon_{i \vec{k} \sigma}$ is the kinetic energy of the
single-particle eigenstate of the $i$th dot having the indicated
degrees of freedom.  For simplicity, we will take these energies
to be independent of dot and channel:
$\epsilon_{i \vec{k} \sigma} = \epsilon_{\vec{k}}$.

The next step in focusing upon a model Hamiltonian is to choose a
form for $t_{\vec{k_{1}} \vec{k_{2}}}$.  Quite generally,
$t_{\vec{k_{1}} \vec{k_{2}}}$ will be nonzero only when both $k_{1}$
and $k_{2}$ lie within some wavevector shell that maximally
spans the space between the theory's low and high momentum
cutoffs.  The size of the wavevector shell depends on details
of the barrier: for a channel with an abrupt delta-function
barrier, the shell spans the largest possible energy range;
for a slowly developing
adiabatic barrier, the shell width will be small on the scale
of a Fermi wavevector.  Important questions are how many
states lie within this shell---i.e., how large is the width $W$
of the corresponding energy shell compared to the average level
spacing $\delta$ between different states in the same channel
(hereafter referred to as ``the average level spacing'' or just
``the level spacing'')---and for a given $\vec{k_{1}}$, for
how many $\vec{k_{2}}$ is $t_{\vec{k_{1}} \vec{k_{2}}}$ nonzero.
Thin-shell models with ``one-to-one'' hopping elements
(i.e., for which $t_{\vec{k_{1}} \vec{k_{2}}} = 0$ unless
$\vec{k_{1}} = \vec{k_{2}}$) have been applied to the coupled
dot problem with some success~\cite{Stafford1,Klimeck,Stafford2},
particulary for level spacings $\delta$ which are on the order of
the charging energy $U$.  For the nearly micron-sized dots used by
Waugh~et~al.~\cite{Waugh}, however, $U$ is
approximately $150 \mbox{ } \mu$eV and $\delta$ is on the order of
$10 \mbox{ } \mu$eV, so we expect that a tunnel-coupling sufficient
to destroy the isolated-dot Coulomb blockade will involve a large
number of single-dot eigenstates.

Consequently, we consider here a thick-shell model which is the
antithesis of the injective thin-shell model.
Working in a regime where $W \gg U \gg \delta$,
we use a tunneling matrix element
$t$ that is independent of $k_{1}$ and $k_{2}$ within the shell:
\begin{displaymath}
t_{\vec{k_{1}} \vec{k_{2}}} = t \mbox{\ \ \ : \ \ \ }
    \forall \mbox{ } \vec{k_{1}}, \vec{k_{2}}
     \mbox{\ \ s.t. \  } \epsilon_{0} < \epsilon_{\vec{k_{1}}},
                           \epsilon_{\vec{k_{2}}} < \epsilon_{0}+W.
\end{displaymath}
As the quantities we calculate are independent of the phase
of $t$, we guiltlessly choose $t$ to be real.
This model is roughly equivalent to one in which each dot is
represented by a tight-binding lattice with intersite hopping
elements of order $W/\delta$ and where interdot tunneling occurs
via a tunneling Hamiltonian with a single site-to-site connection.
Choosing these tunneling sites to be at the origins $\vec{0_1}$
and $\vec{0_2}$ of the respective lattices, we may write
\begin{displaymath}
H_{T} = \sum_{\sigma} (T c^{\dagger}_{2 \vec{0_{2}} \sigma}
                         c_{1 \vec{0_{1}} \sigma}
                      + h.c.),
\end{displaymath}
where $T \equiv N_{W} t$ and $N_{W}=W/\delta$ is the number of
orbital states per channel in each dot within the bandwidth $W$.
(The equivalent lattice model should include second and further
neighbor hopping so that the density of states is approximately
constant between $\epsilon_{0}$ and $\epsilon_{0}+W$.  The
lattice constant is chosen by requiring that the product of $N_{W}$
and the area of a unit cell equals the area of a single dot.)

As the Fermi energy $\epsilon_{F}$ must be somewhere between
$\epsilon_{0}$ and $\epsilon_{0}+W$, the meaning of $\epsilon_{0}$
depends on the width of the band.
For a maximally thick shell, $\epsilon_{0}$ lies at the bottom
of the conduction band, and $W$ is an ultraviolet cutoff, chosen
to be of order twice the Fermi energy.  Alternatively, when
the barrier between the dots has a broader spatial extent,
the energy shell sits more narrowly about the Fermi energy, and
the width $W$ is on the order of the energy difference needed
to produce a factor-of-two change in the magnitude of the transmission
amplitude for an incident particle.  We define a dimensionless
filling parameter
\begin{displaymath}
F \equiv \frac{\epsilon_{F}-\epsilon_{0}}{W},
\end{displaymath}
which gives the position of the Fermi level within the bandwidth
$W$.  Provided that \mbox{$(1-F)W$} and \mbox{$FW$} are both large
compared to $U$, our final results should be independent of the
precise values of $W$ or $F$.

The model we have constructed is basically the two-dot
version of that used by \mbox{L. I. Glazman} and
K.~A.~Matveev~\cite{Matveev1} and H.~Grabert~\cite{Grabert} to study
the charge fluctuations of a single metal particle connected
via point-tunnel junctions to conducting leads.  (See
Figure~1(b).)  Indeed,
the similarity between the two-dot and one-dot problems is
even more fundamental than this observation indicates.
Consider again the double-dot potential energy $V$.  By
transforming to the analog of center-of-mass coordinates,
one generates the following form:
\begin{equation}
V = \frac{U}{4} (\hat{N}_{tot} - \Phi_{tot})^{2} +
            U (\hat{n} - \rho /2)^{2},
\end{equation}
where $\hat{N}_{tot}= {\displaystyle \sum_{i=1}^{2}} \hat{n}_{i}$,
$\Phi_{tot}={\displaystyle \sum_{i=1}^{2}} \phi_{i}$,
$\hat{n}=\frac{\hat{n}_{2} - \hat{n}_{1}}{2}$,
$\rho = \phi_{2} - \phi_{1}$.
The rationale for the normalizations for $\hat{n}$ and
$\rho$ will soon be made apparent.  In the meantime, note
that for our Hamiltonian, $N_{tot}$ is a constant of motion.
Thus, for given $N_{tot}$, $\Phi_{tot}$, and $U$, we can
drop the first term and insert in the Hamiltonian a reduced
potential energy:
\begin{equation}
V_{red} = U (\hat{n} - \rho /2)^{2}.
\end{equation}

Now restrict $N_{tot}$ to be even.  Then, $\hat{n}$ has
integer expectation values in all the unperturbed double-dot
eigenstates.  The Hamiltonian is exactly that of a
single dot tunnel-coupled to an ideal lead.  The dot has number
operator $\hat{n}$, charging energy $2U$, and gate voltage
parameter $\rho/2$.  In the absence of tunneling and with the
level spacing in both dots much less than the charging
energy $U$, the ground state is an eigenstate of $\hat{n}$
that minimizes the reduced potential energy, which in the
future we consider equivalent to ``the potential energy.''
For $\rho=0$, the minimum
potential energy is zero and is achieved when the
eigenvalue $n$ of $\hat{n}$ is zero---i.e., when there
are an equal number of particles in the two dots.  All
other values of $n$ give higher potential energies.  For
$\rho=1$, on the other hand, the minimum potential energy
is $U/4$, and $n=0$ and $n=1$ give degenerate minima.

These no-tunneling distinctions between zero
and $U/4$ and between nondegeneracy and double degeneracy
are quite familiar: they characterized the previously
discussed {\em even} and {\em odd} double-dot ground
states ($\rho=0$ for both).  Indeed, what we
called the ``{\em even} double-dot ground state''
is precisely the ``$N_{tot}$ even, $\rho=0$ ground state.''
The ``{\em odd} double-dot ground state'' is not exactly
the same as the ``$N_{tot}$ even, $\rho=1$ ground state'';
there is no getting around the fact that one case has
one more (or less) particle than the other.  However, in
terms of their ground state energies, the difference
between the two will be down by a factor of
$FN_{W}$ or \mbox{$(1-F)N_{W}$}.  For a wide shell somewhere in the
vicinity of half-filling, both $FN_{W}$ and \mbox{$(1-F)N_{W}$} are
much greater than one, and the above difference is negligible.
Calculation of $E_{int}$ with $\phi_{1}=\phi_{2}$ for the
double dot is therefore effectively equivalent to
calculating the relative shifts of the $\rho=0$ and
$\rho=1$ ground states of a single dot tunnel-coupled to
a bulk two-dimensional electron gas.  More generally,
the value of $\rho$ can be arbitrary, and the difference
in the ground-state energies of the double-dot system for
even $N_{tot}$ and odd $N_{tot}$ is related to the
difference in the ground-state energies of the single-dot
system for gate voltage parameters $\rho$ and $1+\rho$.

We have mapped our double-dot problem onto a
more general single-dot problem for which we can calculate
the relative downward shift of the $\rho \neq 0$ ground
state to the $\rho = 0$ ground state.  As the minimum potential
energy is periodic in $\rho$ with period two and is also even in
$\rho$, we need only
calculate up to $\rho=1$.  Dividing by the
zero-tunneling energy difference of the two ground states,
we find that our emended aim is to calculate
\begin{equation}
f_{\rho} \equiv \left( \frac{E_{int}(\rho)}{U \rho^{2}/4} \right)
     = \Psi_{\rho} (g, N_{ch}, u, N_{W},F),
\end{equation}
where $0 < \rho \leq 1$, $u=U/W$, $N_{W}=W/\delta$, and
$E_{int}(\rho)$ is the ground-state energy relative to the
ground-state energy for $\rho=0$.

\subsection{Barrier Conductance in the Weak-Coupling Limit}

Before we can derive our equation for $f_{\rho}$ in terms of $g$,
we must find a formula for the barrier conductance.
Measurement of the barrier conductance $G_{b}$ with the exterior
gates turned off can be modeled by calculating the tunnel junction
conductance for \mbox{$U=0$}.  As mentioned before, we assume the different
conducting channels to be identical yet independent---their
individual conductances are the same and they
do not interfere with one another.  These assumptions are certainly
reasonable for the two spin channels in the
experiment of Ref.~\cite{Waugh}.
Using the Lippmann-Schwinger equation with
$H_{T}$ inserted for the scattering potential~\cite{Roman}, one can
solve for the perturbed electron eigenfunctions.  The Heisenberg
equation of motion for $\hat{n}_{1}$ can then be used to solve for the
particle flow from dot 1 to dot 2 for a given voltage bias.  Solving
the resulting expression for the linear conductance gives the
following equation for the dimensionless conductance per channel:
\begin{equation}
g = \frac{G_{b}}{N_{ch} G_{0}} = \frac{4 \alpha}{|1 + \chi \alpha|^{2}},
\end{equation}
where $\alpha = (\pi T / W)^{2} = (\pi t / \delta)^{2}$ and
$\chi = (1 + \frac{\em i}{\pi} \log(\frac{F}{1-F}))^{2}$.
H. O. Frota and K. Flensberg have derived this equation for
half-filling ($F = 0.5, \chi = 1$) via
a Green's function-Kubo formula approach~\cite{Frota}.

The calculated conductance $G_{b}$ exhibits rather curious
behavior: it first rises to a maximum of $N_{ch} G_{0}$
corresponding to $N_{ch}$ fully open channels (It does this even
when $Im[\chi] \neq 0$.), and then falls asymptotically to zero as
$(T/W = t / \delta) \rightarrow \infty$.  As Frota
and Flensberg note~\cite{Frota}, the asymptotic damping of the conductance
results from the fact that formation of bonding and anti-bonding
states at the tunnel junction makes the cost of passing
through prohibitively high.  The limit of $(T/W = t / \delta)
\rightarrow \infty$ is in some sense unphysical: we do not
expect a point-to-point hopping coefficient $T$ to significantly
exceed the tunneling shell width; nor do we expect the tunneling
matrix element $t$ to be much greater than the average level
spacing.  Nevertheless, the apparent absence of any good reason
to truncate the theory at a particular value of $t$ leaves
us with a model that is at best unwieldy in the limit of strong
coupling.  To get the correct limiting behavior for strong
coupling, it is more convenient to use a different approach,
suitable for perturbation
about the $g=1$ limit.  This will be described in Section IV.

\subsection{Relative Energy Shift of {\em Even} and
  {\em Odd} States in the Weak-Coupling Limit}

In the meantime, the site-to-site tunneling model is still
useful in the weak coupling regime.  So we
plod ahead, calculating via standard Rayleigh-Schr\"{o}dinger
perturbation theory the second order shift in the ground
state energy for $\rho \neq 0$ minus that for $\rho = 0$.
The $\rho=1$ shift will be taken to equal the limit of the
general $0 \leq \rho < 1$ shift as $\rho \rightarrow 1$.
It might be objected---correctly---
that this limit fails properly to account for the degeneracy
of the ground state at $\rho = 1$.  Such a failing is pardonable,
however, for the contributions that are left out are all smaller
by a factor of $FN_{W}$ or \mbox{$(1-F)N_{W}$} from those which are
retained.  Since we assume that $t/\delta$ is finite, $F$ is of
order~$\frac{1}{2}$, and $N_{W}$ is large, the omitted terms
are negligible.

For $N_{W} \gg 1$, the perturbation theory sums can be approximated
as integrals.  Observing that $u \equiv \frac{U}{W} \ll 1$,
we divide the difference between the second order shifts by
$U \rho^{2}/4$ to get the leading approximation to $f_{\rho}$:
\begin{equation}
f^{(1)}_{\rho} = 4 N_{ch} \frac{t^{2}}{\delta^{2}}
[(1-\rho) \log (1-\rho) + (1+\rho) \log (1+\rho)
 + O(u \rho^{2})]/ \rho^{2}.
\end{equation}
The second-order term indicates a significant feature of $f_{\rho}$:
it is even in $\rho$.  This property has been noted by
H. Grabert~\cite{Grabert} and results from the fact
that at any order of perturbation theory, every tunneling
process contributing to the energy shift has a twin with
the roles of dots 1 and 2 interchanged.  In any
intermediate virtual state with eigenvalue $n$ for
$\hat{n}$, the potential energy is greater than that for
the unperturbed ground state by
$\delta V(\rho) = Un(n-\rho)$.
Therefore, when dots 1 and 2 are interchanged,
$\delta V(\rho) \rightarrow \delta V(-\rho)$
for all the intermediate states.
If we represent one of the twin terms by $\Delta(\rho)$,
the other is $\Delta(-\rho)$, and we see that $f_{\rho}$ is
constructed of sums that are even in $\rho$.

Using the second-order (in $t/\delta$) parts of $g$ and $f_{\rho}$,
we can now write a first-order equation for $f_{\rho}$ in terms of $g$:
\begin{equation}
f^{(1)}_{\rho} = \frac{N_{ch}g}{\pi^{2}}
  \frac{[(1-\rho) \log (1-\rho) + (1+\rho) \log (1+\rho)
       + O(u \rho^{2})]}{\rho^{2}},
\end{equation}
a result consistent with the large-$N_{ch}$ calculation of the
effective capacitance of a single dot at $\rho=0$~\cite{Golubev}.
Setting $\rho=1$ to calculate the relative shifts of the original
{\em even} and {\em odd} states, we find
\begin{equation}
f^{(1)} =  \frac{2 \log 2}{\pi^{2}} N_{ch} g + O(u g, g^{2}),
\label{eq:fweak}
\end{equation}
where we have used the fact that $f$ as originally defined
without the subscript is equivalent in our limits to $f_{\rho=1}$.
The above equation indicates that the suggestion that the
plot of $f$ is a displaced version of the plot for $g$ is
not precisely correct.  In particular, for $g \ll 1$ and
$N_{ch}=2$, Equation~(\ref{eq:fweak}) gives a slope of approximately
0.28 for $f(g)$, rather than unity.  Thus, in this regime, the
fractional splitting $f$ of the double-dot conductance peaks should
lag $g$, the dimensionless barrier conductance per channel.

\section{Connection to the Strong-Coupling Limit}

If we blithely extended our perturbative equation for $f$
to the limit $g \rightarrow 1$, the
large-$N_{ch}$ $f$ would greatly overshoot its mark and the
one or two-channel $f$ would fall substantially short.  The
real issue is not, however, how badly such a naive
extrapolation fails, but whether we can interpolate between
these weak-tunneling results and those which can be calculated
for the strong-tunneling limit.
For the \mbox{large-$N_{ch}$} limit, a reasonable
interpolation between the solutions for weak and strong
coupling has already been found~\cite{Golubev,Panyukov,Zimanyi}.
The situation is less clear when $N_{ch}$ equals one or two.
Flensberg and Matveev~\cite{Flensberg,Matveev2} have
proposed a useful L\"{u}ttinger-liquid approach in which
the nearly transparent link between a dot and an electrode
is modeled as a one-dimensional channel with a
slightly reflective potential barrier.  Convergence to
the single composite-dot limit is achieved naturally and
neatly, and $E_{int}$ is calculated perturbatively
in $r$, where $r$ is the reflection amplitude, and
$g=1-|r|^{2}$.  Translating Matveev's calculations of the
leading term for $(1-g) \ll 1$ into our language, we find
that for $N_{ch}=1$ (i.e., assuming spin polarization),
\begin{equation}
f = 1 - \frac{8 e^{\gamma}}{\pi^{2}} \sqrt{1-g},
\label{eq:fstr1}
\end{equation}
where $\gamma \simeq 0.577$ is the Euler-Mascheroni
constant.  For the case relevant to the experiment of
Ref.~\cite{Waugh}, $N_{ch}=2$, Matveev's calculation gives
\begin{equation}
f = 1 + \frac{16 e^{\gamma}}{\pi^{3}} (1-g) \log (1-g).
\label{eq:fstr2}
\end{equation}
Except for the logarithmic factor in the second formula,
these equations are of the form suggested by the scaling
analysis of Flensberg~\cite{Flensberg}, which predicts
effective charging energies behaving as $(1-g)^{N_{ch}/2}$.
Matveev's initial two-channel solution is, in fact, linear
in $(1-g)$ but diverges logarithmically as
$U/\delta \rightarrow \infty$.  A higher-order analysis
to eliminate the divergence replaces the
logarithm having argument $U/\delta$ with one having argument
$(1-g)^{-1}$~\cite{Matveev2}.

In Figure~3, we show the $f$-versus-$g$ plots given by the
weak and strong coupling formulas
\mbox{(\ref{eq:fweak}), (\ref{eq:fstr1}), and (\ref{eq:fstr2})}
for $N_{ch}=1$ and $N_{ch}=2$.  In each case, a plausible
interpolation between the weak and strong coupling limits is
given by a dashed curve.  The experimental data of
Ref.~\cite{Waugh} are in reasonable agreement with
the dashed curve for $N_{ch}=2$.  Nevertheless, from a
theoretical standpoint, it is clear that, unlike the calculations
for $N_{ch} \gg 1$, for $N_{ch}=2$ the order of calculation
completed so far does not really allow confident
interpolation between the weak and
strong coupling limits.  Calculation of higher orders in
perturbation theory should improve the matching, but such
computations are made difficult by the fact that the correlations
induced by the strong Coulomb interaction make normal
Green's functions methods inapplicable~\cite{Schoeller}.
Different time orders must be treated separately, and as
appears to occur quite generally in Coulomb-blockade
problems~\cite{Averin2}, the number of diagrams grows
pathologically with the order in perturbation theory.
Nevertheless, calculation of the \mbox{$g^{2}$-term} in
the weak-tunneling limit is within reach, and this term
may suffice to give a more reliable interpolation between the
weak and strong coupling regimes.

\section{Conclusion}

Following the work of Waugh~et~al.~\cite{Waugh}, we
have investigated the relation between the barrier
conductance and the Coulomb blockade for two
electrostatically equivalent dots connected by one or
more identical tunneling channels.  We propose
to write the fractional peak splitting $f$ of the Coulomb
blockade conductance peaks as a function of the number of
channels $N_{ch}$ and the dimensionless barrier conductance
per channel $g$, assuming that the energy level spacing
$\delta$ is small compared to the Coulomb blockade energy
$U$ and that $U$ is small compared to the bandwidth $W$
of states over which the amplitudes for transmission
through the barrier are roughly constant.  Using a ``uniform
thick-shell model'' for the tunneling term in the Hamiltonian,
we solve for this function to leading order in the limit of
weak interdot coupling.  We find that the
peak splitting should evolve substantially more slowly
than the barrier conductance in this limit.  Having noted
that the two-dot problem can be mapped
onto a better known one-dot problem, we adapt previous
strong-coupling results to obtain the asymptotic form of
the double-dot peak-splitting in the limit $g \rightarrow 1$.
For the case of $N_{ch}=2$, which is pertinent to the experimental
results of Ref.~\cite{Waugh}, the limiting forms for the
strong and weak coupling do not match up well enough to
allow a reliable quantitative interpolation between the
two limits.  Nevertheless, a visually plausible interpolating
curve is in good qualitative agreement with existing
experimental data.

\section{Acknowledgments}

The authors are grateful for very helpful discussions with
F.~R.~Waugh, R.~M.~Westervelt, Gergely~T.~Zimanyi,
C.~A.~Stafford, C.~Crouch, C.~Livermore, and Steven~H.~Simon.
This work was supported in part by the United States Air Force
through a graduate fellowship for John M. Golden and by the
NSF through the Harvard Materials Research Science and
Engineering Center, Grant No. DMR94-00396.

After this manuscript was essentially completed, the authors
received a preprint by K.~A.~Matveev, L.~I.~Glazman, and
H.~U.~Baranger~\cite{Matveev3} in
which similar ideas were independently developed.

\begin{figure}
\caption{(a) Schematic diagram for the double dot.  Negative
potentials are applied to each of the gates to form the
double-dot structure.  The gate potentials $V_{g1}$ and $V_{g2}$
control the average numbers of electrons on the dots.  These
are the potentials that are varied to see the Coulomb blockade.
$V_{b}$ controls the rate of tunneling between the dots.
$V_{x1}$ and $V_{x2}$ control the rate of tunneling to
the adjacent bulk 2D electron gas (2DEG) leads.
For calculations of the double-dot energy shifts, tunneling to
the leads is assumed negligible compared to tunneling between
the two dots.  In measuring the barrier conductance $G_{b}$,
however, the potentials $V_{xi}$ are
turned off so that each dot is strongly connected to its lead.
The side-wall potentials $V_{s1}$ and $V_{s2}$ are
fixed.
(b) Schematic diagram for the single dot. $V_{b}$ now controls
tunneling between the dot and the bulk 2DEG.  $V_{g}$ determines
the average number of electrons on the dot.  For our purposes,
$V_{s}$ and $V_{x}$ are constant, and tunneling to the bulk 2DEG
through the barrier defined by $V_{x}$ is negligible compared
to tunneling through the barrier defined by $V_{b}$.}
\label{fig:1}
\end{figure}

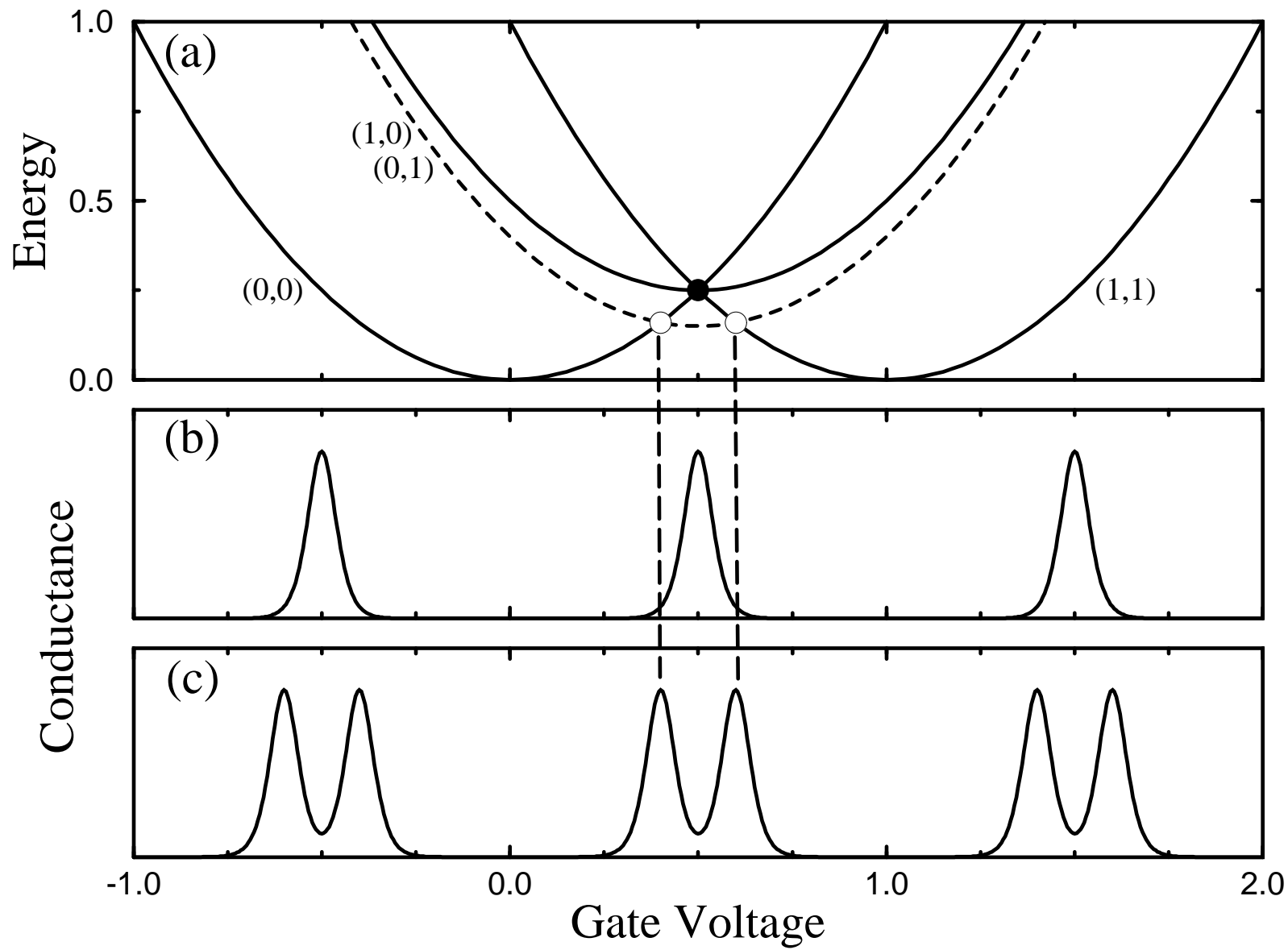
\begin{figure}
\caption{(a) Energy curves in the capacitive charging model for
electrostatically identical dots with $V_{g1}=V_{g2}$.
Energies are given in units of the charging
energy $U$; the gate voltage is given in units of $\frac{e}{C_{g}}$.
Each zero-coupling eigenstate with definite particle number $n_{i}$
on the $i$th dot gives rise to a parabola,
labeled $(n_{1},n_{2})$, which shows the state's energy
as a function of the gate voltage.  The zero of energy is chosen to
coincide with the lowest energy possible for states with an even value
for the total number of particles $N_{tot}$.  The solid odd-$N_{tot}$
parabola gives the lowest-energy curve only when there is no
interdot coupling.  The dotted parabola is the shifted-down
energy curve for odd $N_{tot}$ that results from
finite coupling between the dots.  The relevant degeneracy points
are indicated by a black dot for zero coupling and
white dots for finite coupling.
(b) ``Zero-coupling'' conductance through the double dot
as a function of the gate voltage.
(For ease of viewing, peaks are
depicted as symmetric with uniform finite widths and heights.)
Conductance peaks are
aligned with the zero-coupling degeneracy points such as the
one shown in (a) and
occur regularly with unit period.
(c) Conductance through the dot for finite interdot coupling.
Conductance peaks are aligned with the perturbed degeneracy
points.  Each zero-coupling peak has split into
two separate peaks, equally distant from the zero-coupling
peak position.  Increasing the interdot coupling increases the
separation between the paired peaks until the full set of
peaks is again regularly distributed, with half the original
period. (This figure for the capacitive charging model follows
that of Ref.~[2].) }
\label{fig:2}
\end{figure}

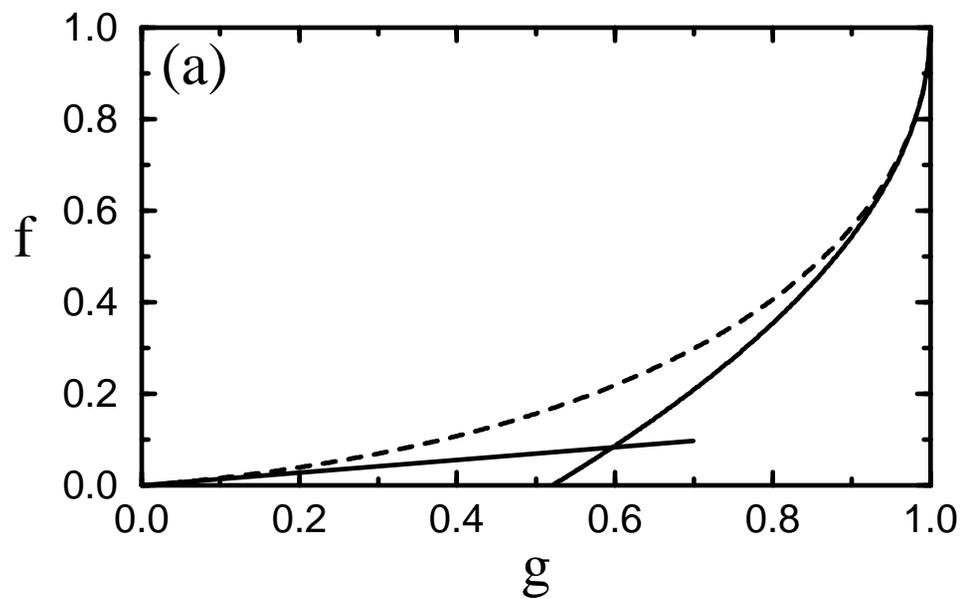
\begin{figure}
  \caption{Graphs of the fractional Coulomb blockade conductance
peak splitting $f$ as
a function of the dimensionless conductance per channel $g$
in the weak and strong tunneling limits for \mbox{(a) $N_{ch}=1$}
and \mbox{(b) $N_{ch}=2$}.  Possible interpolating functions are
shown by dashed curves.}
  \label{fig:3}
\end{figure}



\begin{references}

\bibitem{Review} For an introduction to ``single-electronics,'' see
  M. A. Kastner, Rev. Mod. Phys. {\bf 64}, 849 (1992); D. V. Averin and
  K. K. Likharev, in {\it Mesoscopic Phenomena in Solids}, edited by
  B. Altshuler~et~al.\ (Elsevier, Amsterdam, 1991); and several
  articles in {\it Single Charge Tunneling}, edited by H. Grabert
  and M. H. Devoret, NATO ASI Series B, vol. 294 (Plenum, New
  York, 1992).

\bibitem{Waugh} F. R. Waugh, M. J. Berry, D. J. Mar, R. M. Westervelt,
  K. C. Campman, and A. C. Gossard, preprint (9/94); see also F. R. Waugh,
  Ph.D. thesis, Harvard University, 1994.

\bibitem{Molen} L. W. Molenkamp, Karsten Flensberg, and M. Kemerink,
  preprint (3/95, cond-mat/9503135).

\bibitem{Ruzin} I. M. Ruzin, V. Chandrasekhar, E. I. Levin, and
  L. I. Glazman, Phys. Rev. B {\bf 45}, 13~469 (1992); L. I. Glazman
  and V. Chandrasekhar, Europhys. Lett. {\bf 19}, 623 (1992).

\bibitem{Golubev} D. S. Golubev and A. D. Zaikin, Phys. Rev. B
  {\bf 50}, 8736 (1994).

\bibitem{Panyukov} S. V. Panyukov and A. D. Zaikin, Phys. Rev. Lett
  {\bf 67}, 3168 (1991); Physics Letters A {\bf 183}, 115 (1993).

\bibitem{Zaikin} A. D. Zaikin, D. S. Golubev, and S. V. Panyukov,
  Physica B {\bf 203}, 417 (1994).

\bibitem{Zimanyi} G. Falci, J. Heins, Gerd Sch\"{o}n, and
  Gergely T. Zimanyi, Physica B {\bf 203}, 409 (1994);
  G. Falci, Gerd Sch\"{o}n, and Gergely T. Zimanyi, Phys. Rev.
  Lett. {\bf 74}, 3257 (1995).

\bibitem{Schoeller} Herbert Schoeller and Gerd Sch\"{o}n, Phys.
  Rev. B {\bf 50}, 18~436 (1994).

\bibitem{Roman} For accounts of the formal theory of scattering, see,
  for example, Paul Roman's {\it Advanced Quantum Theory} (Addison-Wesley,
  Reading, MA, 1965) or Michael D. Scadron's {\it Advanced Quantum
  Theory} (Springer-Verlag, Berlin, 1991).

\bibitem{Shekhter}  R. I. Shekhter, Zh. Eksp. Teor. Fiz. {\bf 63}, 1410
  (1972) [Sov. Phys. JETP {\bf 36}, 747 (1973)]; I. O. Kulik and R. I.
  Shekhter, Zh. Eksp. Teor. Fiz. {\bf 68}, 623 (1975) [Sov. Phys. JETP
  {\bf 41}, 308 (1975)].

\bibitem{Averin1} D. V. Averin and K. K. Likharev, Zh. Eksp. Teor.
  Fiz. {\bf 90}, 733 (1986) [Sov. Phys. JETP {\bf 63}, 427 (1986)];
  J. Low Temp. Phys. {\bf 62}, 345 (1986).

\bibitem{Stafford1} C. A. Stafford and S. Das Sarma, Phys. Rev. Lett.
  {\bf 72}, 3590 (1994).

\bibitem{Klimeck} G. Klimeck, Guanlong Chen, and S. Datta,
  Phys. Rev. B {\bf 50}, 2316 (1994); Guanlong Chen, G. Klimeck,
  S. Datta, Guanha Chen, and W. A. Goddard III, Phys. Rev. B
  {\bf 50}, 8035 (1994).

\bibitem{Stafford2} C. A. Stafford and S. Das Sarma, preprint (1995).

\bibitem{Matveev1} L. I. Glazman and K. A. Matveev, Zh. Eksp. Teor.
  Fiz. {\bf 98}, 1834 (1990) [Sov. Phys. JETP {\bf 71}, 1031 (1990)];
  K. A. Matveev, Zh. Eksp. Teor. Fiz. {\bf 99}, 1598 (1991) [Sov.
  Phys. JETP {\bf 72}, 892 (1991)]

\bibitem{Grabert} Hermann Grabert, Phys. Rev. B {\bf 50}, 17~364 (1994).

\bibitem{Frota} H. O. Frota and Karsten Flensberg, Phys. Rev. B
  {\bf 46}, 15~207 (1992).

\bibitem{Flensberg} Karsten Flensberg, Physica B {\bf 203}, 432 (1994);
  Phys. Rev. B {\bf 48}, 11~156 (1993).

\bibitem{Matveev2} K. A. Matveev, Phys. Rev. B {\bf 51}, 1743 (1995).

\bibitem{Averin2} D. V. Averin and Yu V. Nazarov, in {\it Single Charge
  Tunneling} (1992).

\bibitem{Matveev3} K. A. Matveev, L. I. Glazman, and H. U. Baranger,
preprint (4/25/95, cond-mat/9504099).

\end{references}
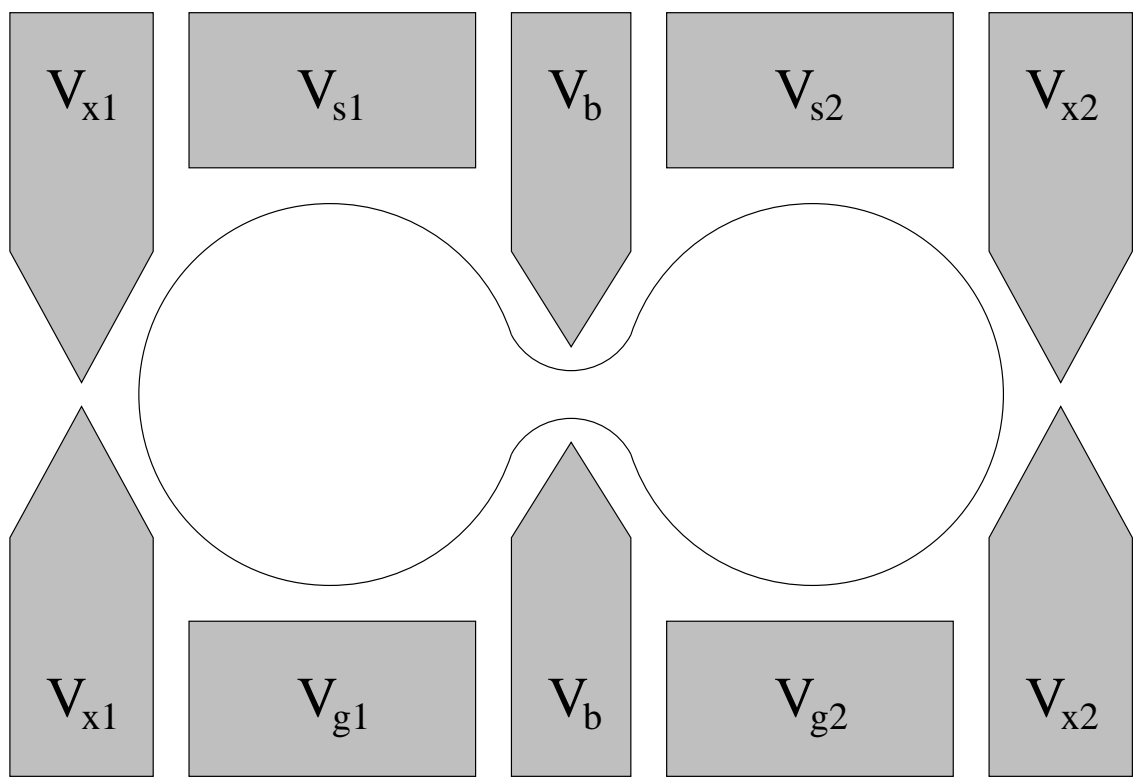
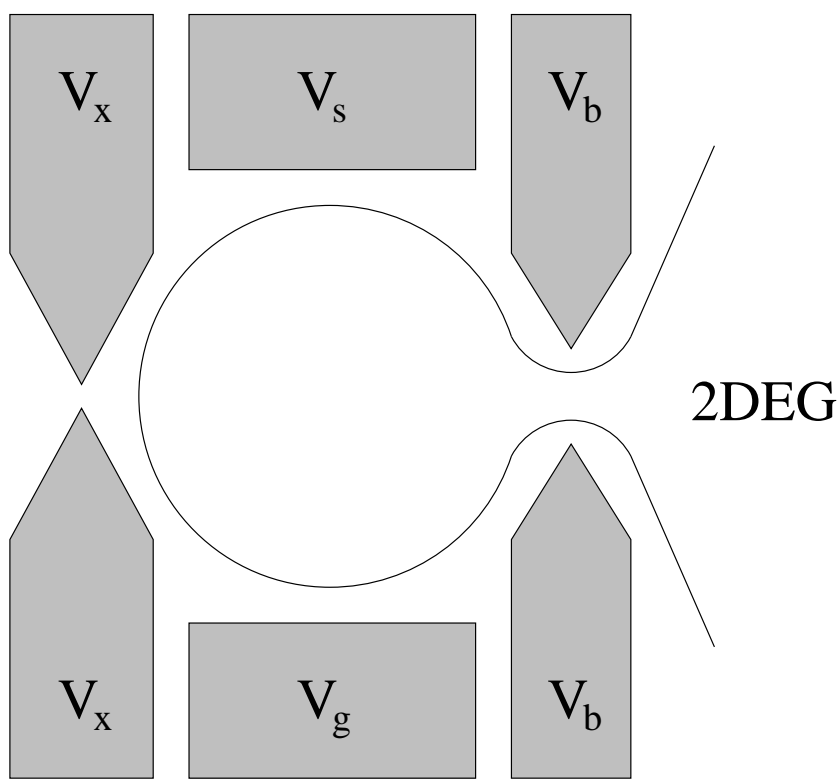
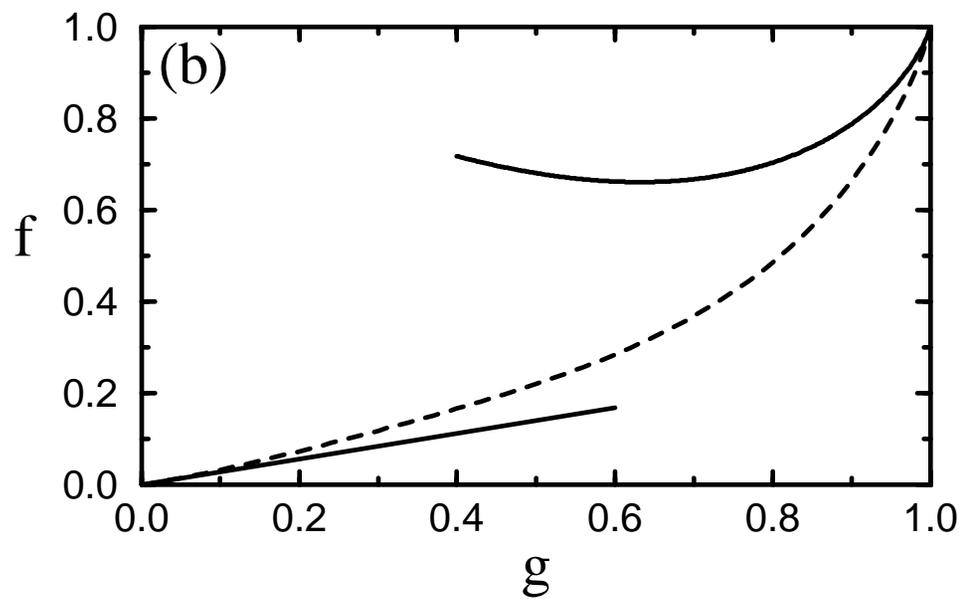
\end{document}